\documentclass[12pt,twoside]{article}
\usepackage[mathscr]{eucal}
\usepackage{amsmath,amsfonts,amssymb,amsthm}
\usepackage{hyperref}
\usepackage{times}

\def\mt1{Metsaev:1998it}

\def\ci{\cite}
\newcommand{\be}{\begin{equation}}
\newcommand{\ee}{\end{equation}}

\def \m {\mu}

\def \bi{\bibitem}
\def \la {\label}

\def \l {\lambda}
\def\foot{\footnote}

\def \adss {$AdS_5 \times S^5$\ }
\newcommand{\rf}[1]{(\ref{#1})}
\def \ov {\over}

\def \ha{{1\ov 2}}

\def \del {\partial}

 \def \bb {\bar \begin{equation}ta}

\def \bi{\bibitem}
\def \la {\label}

\def \l {\lambda}
\def\foot{\footnote}

\def \adss {$AdS_5 \times S^5$\ }

\def \ov {\over}

\def \varpi {{\rm w}}

\def \rt {{\rm t}}
\def \no {\nonumber }

\def \adss {$AdS_5 \times S^5$\ }

\def \bb {{\bar  \begin{equation}ta}}

\def \inv {^{-1}}

\def \dpp {\del_+}
\def \dmm {\del_-}

\def \ci {\cite}

\def\tr{{\rm Tr}}
\def\str{{\rm STr}}

\newcommand{\Psr}{\Psi_{{}_1}}
\newcommand{\Psl}{\Psi_{{}_2}}

\newcommand{\psr}{\psi_{{}_1}}
\newcommand{\psl}{\psi_{{}_2}}

\renewcommand{\hat}{\widehat}

\renewcommand{\simeq}{\cong}

\newcommand{\bref}[1]{\textbf{\ref{#1}}}

\newcommand{\im}{\mathop{\mathrm{Im}}}

\newcommand{\algm}{\Liealg{m}}
\newcommand{\algf}{\Liealg{f}}
\newcommand{\alghf}{\hat\algf}
\newcommand{\algg}{\Liealg{g}}
\newcommand{\algh}{\Liealg{h}}

\newcommand{\algp}{\Liealg{p}}

\newcommand{\Liealg}{\mathfrak}       

\newcommand{\id}{\mathbf{1}}
\renewcommand{\imath}{i}

\renewcommand{\d}{\partial}

\renewcommand{\geq}{\,{\geqslant}\,}
\renewcommand{\leq}{\,{\leqslant}\,}

\newcommand{\binner}[2]{%
  {\langle}\kern-4.15pt{\langle}#1{,}\,#2{\rangle}\kern-4.15pt{\rangle}}
\newcommand{\commut}[2]{[#1{,}\,#2]}

\newcommand{\scommut}[2]{\{#1{,}\,#2\}}

\newcommand{\half}{\mathchoice{%
    \ffrac{1}{2}}{\frac{1}{2}}{\frac{1}{2}}{\frac{1}{2}}}

\newcommand{\ffrac}[2]{\raisebox{.5pt}%
  {\footnotesize$\displaystyle\frac{#1}{#2}$}\kern1pt}

\newcommand{\fC}{\mathbb{C}}

\newcommand{\fR}{\mathbb{R}}

\def \vp {\varphi}

\numberwithin{equation}{section} \makeatletter
\@addtoreset{equation}{section}

\vspace{-1cm}

  \def \vp {\varphi}
  \def \a {\alpha}

\def\km{\mu}

\newcommand{\pp}{{\parallel}}
\newcommand{\oo}{{\perp}}
\def \A {{\cal A}}

\def \s {\sigma}

\newcommand{\bea}{\begin{eqnarray}}
\newcommand{\eea}{\end{eqnarray}}

\def \bp {\begin{pmatrix}}  \def \ep {\end{pmatrix}}

\begin{document}

\vspace{ -3cm}
\rightline{Imperial-TP-AT-2008-2}

\begin{center}
\vspace{0.2cm}
{\Large\bf On reduced models for superstrings on $AdS_n\times S^n$
\vspace{0.4cm}
   }

 \vspace{0.5cm} {M. Grigoriev$^{a,}$\footnote{grig@lpi.ru}
 and A.A. Tseytlin$^{b,a,}$\footnote{tseytlin@imperial.ac.uk
 }}\\
 \vskip 0.13cm
{\em
$^{a}$  Tamm Theory Department, Lebedev Physical Institute, Leninsky 53, Moscow 119991, Russia\\
$^{b}$   Blackett Laboratory, Imperial College,
London SW7 2AZ, U.K.    
   }

\end{center}

 \vskip 0.15cm
 
 \begin{abstract}
We review  the Pohlmeyer  reduction  procedure  of the superstring sigma model
on $AdS_n\times S^n$ leading to a gauged WZW model with an integrable
 potential coupled to 2d fermions. In particular, we consider  the case of the 
Green-Schwarz  superstring on $AdS_3\times S^3$  supported by RR flux.
The bosonic part of the reduced model  is given by the
sum of the complex sine-Gordon Lagrangian and its  sinh-Gordon   counterpart.
We determine the corresponding fermionic part and discuss 
possible existence of hidden  2d supersymmetry  in  the reduced action. 
We also  elaborate on   some  general  aspects of the 
Pohlmeyer reduction applied to the $AdS_5\times S^5$  superstring.
\end{abstract}

\thispagestyle{empty}
\setcounter{page}{0}

 \vskip 0.15cm

\section{Introduction}

Further progress in understanding AdS/CFT correspondence requires solving the
 superstring theory on $AdS_5\times S^5$.
Being an essentially nonlinear theory (IIB Green-Schwarz superstring on 
$ \frac{PSU(2,2|4)}{ SO(1,4) \times SO(5)}$ supercoset \ci{mt})
 this theory is difficult to quantize directly.
 By  analogy 
with the flat space GS superstring one can
try to utilize an appropriate version of a light-cone gauge, 
but that does not simplify the action   and, in contrast to the flat space case, 
 breaks 2d Lorentz invariance.
The  lack of 2d Lorentz invariance  makes 
 it hard to apply  directly the 
 known results and methods of 2d integrable field theory. 
 In particular, the $S$-matrix of scattering  of string fluctuations 
 in a light-cone gauge is not  2d Lorentz invariant
 and  constraints on it are a priori unclear.

An alternative approach \ci{GT,ms}
 is to use a version of the Pohlmeyer ``reduction'' \ci{pol}
 which allows one to reformulate the theory in terms of physical degrees
  of freedom only. 
  It is based on writing the equations of motion in terms of the coset  currents, 
  solving explicitly  the Virasoro constraints by introducing a new set of fundamental
  variables algebraically related to the currents
  and then reconstructing the action for the new independent variables. 
 Remarkable features of the Pohlmeyer-reformulated 
  theory for the GS   $AdS_5\times S^5$ model
are the explicit 2d Lorentz invariance and the
standard  kinetic term for the fermions. 
As in the purely bosonic case \ci{bak}, 
the \adss 
 Pohlmeyer reduction preserves the integrable structure -- the reduced theory is 
an   integrable deformation of a  gauged WZW model by an extra potential term, i.e. 
a special case of  non-abelian Toda theory. In addition, it contains 
 fermionic terms   and  thus resemble
 a 2d supersymmetric   generalization of the gauged WZW model.
In an  appropriate free-theory  limit the reduced action  coincides with the 
pp-wave action for 8+8 massive degrees  of freedom 
\ci{met} (which in turn generalizes the  flat space light cone gauge action). 

The hope is that this  reduced theory 
 for  the $AdS_5\times S^5$ superstring  should be the starting point for its quantization. 
 There are still 
 a number of open problems  at the classical   level 
 (the  interpretation of conserved charges, 
  choice of  vacuum, 
 fixing the residual gauge symmetry, existence of  world sheet supersymmetry,  etc.)
  remain to be explored further. 
 This suggests  to  study
  first  simpler low-dimensional   analogs, i.e.  $AdS_n\times S^n$ 
 GS models with  $n=2,3$.  In   the $AdS_2\times S^2$ case 
  the reduced theory happens  to be very simple and can be identified 
  with the $N=2$  2d supersymmetric extension of the sine-Gordon model \ci{GT}.

Here we shall address the  next non-trivial case of the $AdS_3\times S^3$ 
superstring. 
 The corresponding  GS superstring action  \cite{Ram-Raj,adst}
 is slightly  different in the structure  from that in the  $AdS_5\times S^5$
 and $AdS_2\times S^2$ cases. As a result, the reduction scheme used in~\cite{GT} 
 requires some modification. 
Since the bosonic part of the  $AdS_3\times S^3$ sigma model is 
a principal chiral model defined on  the group space $G=SU(1,1)\times SU(2)$,  
this 
requires to understand how to do the Pohlmeyer reduction in the case where
the target space is a group manifold.

The Pohlmeyer reduction of the $F/G$ coset
 sigma model is based on using the  $G$ gauge symmetry. 
 One can formally describe  the principal chiral model also 
 as a coset one by representing $G$ as a  symmetric space $G\times G /G$ 
 where 
the denominator subgroup is embedded  diagonally (see also   \cite{miramontes}).
 The   Pohlmeyer reduced theory  can then be identified with the $G/H$
gauged WZW model with a potential, with $H$ being a subgroup corresponding to 
the Cartan subalgebra $\algh$ of $\algg$.

Below we  shall show how this procedure can be applied to the  GS 
superstring on $AdS_3 \times S^3$. 
Compared to the $AdS_5\times 
S^5$  case  in~\cite{GT} the   only nontrivial ingredient is the explicit realization of the
 $Z_4$ grading of the $psu(1,1|2)\oplus psu(1,1|2)$ superalgebra. 

In Section  \bref{sec:2}  we shall give an algebraic 
construction of the Pohlmeyer reduction
for a principal chiral model. 

In Section \bref{sec:3}  we shall explicitly identify the $Z_4$
 grading on $psu(1,1|2)\oplus psu(1,1|2)$
superalgebra  and perform the Pohlmeyer reduction of 
the corresponding  superstring sigma model.
The main motivation is to see if the resulting reduced sigma model 
has $N=2$   2d supersymmetry as we  found earlier in the  
 $AdS_2 \times S^2$   case. The conjectured presence of world sheet supersymmetry 
 in the $AdS_3 \times S^3$  and also $AdS_5 \times S^5$  cases 
 would be quite surprising  since it is absent in  the original  Green-Schwarz 
 action  in which fermions are 2d scalars    and have an unusual kinetic term.
 Unfortunately, the presence of 2d  supersymmetry  is not apparent 
 in the reduced $AdS_3 \times S^3$ action we derive below.


 In Section \bref{sec:4} we shall  make some general comments on the reduced model: its 
 relation to original model, conserved charges, vacuum configuration and perturbative 
 expansion near it.


\section{Pohlmeyer reduction for strings on a group manifold}\label{sec:2}

The principal chiral model (PCM) for a simple group  $G$ 
can be represented as a  coset sigma model for 
\begin{equation}
\label{homs}
 \frac{F}{\bar G}=\frac{G\times G}{{\bar G}}\,,
\end{equation}
where $\bar G\simeq G$ is a subgroup of $G\times G$.
In general, we   can  represent elements of
$F=G\times G$ as pairs $(g_1,g_2)$. The denominator subgroup $\bar G$
is chosen to be the twisted diagonal subgroup, 
 i.e. the subgroup of  $(g,\hat\chi(g))$,
where $\hat\chi$ is an automorphism of $G$  compatible with the   invariant bilinear  form  $\tr$
on  Lie algebra $\algg$.\footnote{Although the reduced theory does not depend on 
$\hat \chi$ in the bosonic 
case, the formulation of GS supercoset sigma model
requires nontrivial $\hat\chi$.
That is why we  keep here $\hat\chi$ for generality.}
The standard example is when  $\hat\chi$
is an  identity so that $\bar G$ is  embedded diagonally.
Another useful choice of $\hat\chi$ is    when   $G$ is defined in a matrix representation 
 so that the transposition $~^t$ is an anti-automorphism of $G$ (i.e. 
$a^t$ belongs to $G$ for any $a\in G$ and $(ab)^t=b^t a^t$): then 
 one can set  $\hat\chi(a)=(a^t)^{-1}$. 


Let the pair  $(a,b)$  with $a,b\in \algg$  denote  an  element of  Lie algebra $\algf$ of $F$.
 The invariant bilinear form on $\algg$ induces that on $\algf=\algg\oplus \algg$.
 Then  subalgebra $\bar\algg\subset \algf$   which is the Lie algebra of   $\bar G$ 
 is isomorphic to $\algg$ and is formed by  $(a,\chi(a))$ where $\chi$
is the Lie algebra automorphism induced by $\hat\chi$.
 Because $\chi$ is compatible with the $\tr$, i.e. $\tr(\chi(a)\chi(b))=\tr(ab)$,
  the orthogonal complement $\algp$ of $\bar\algg $ in $\algf$ is
formed by elements  $(a,-\chi(a))$. Homogeneous space~\eqref{homs} is,  in fact,  a symmetric
 space:
\begin{equation}
\commut{\bar\algg}{\bar\algg}\subset \bar\algg\,,\quad \commut{\bar\algg}{\algp}\subset \algp\,,
\quad 
\commut{\algp}{\algp}\subset \bar\algg\,,
\qquad\quad  \,\algf=\bar\algg\oplus\algp\,\,.
\end{equation}
In particular, in the case where $\hat\chi(g)=(g^t)^{-1}$ the 
corresponding  Lie algebra automorphism is $\chi(a)=-a^t$; the subalgebra $\bar\algg$ is then formed by 
 $(a,-a^t)$ while $\algp$ is formed by 
$(a,a^t)$.

The $F/{\bar G}$ coset sigma model is defined by the Lagrangian ($f \in F$) 
\begin{equation}
 L
 =- {\textstyle \frac{1}{2}} \tr(P_a P^a)\,, \qquad P_a=(f^{-1}\d_a f)_\algp\,.
\end{equation}
In the above case  of \rf{homs} 
it  is equivalent to the standard principal chiral field model. Indeed,
using the gauge freedom one can always set
 $(g_1,g_2)= (g,\id)$. In this gauge 
  the above  Lagrangian
becomes 
\begin{equation}
 L=-{\textstyle \frac{1}{8}}\tr(g^{-1}\d_a g\ g^{-1}\d^a g) \ . 
\end{equation}
Having identified the PCM as a special 
coset model one can attempt to perform its Pohlmeyer reduction.
One option is to treat it as a classical  2d field theory
 and  use the conformal symmetry to fix the components of the stress tensor 
$T_{++}=\mu^2,\,\, T_{--}=\mu^2$. Another one is to consider strings on 
$G\times \fR_t$; then 
the conditions
$T_{++}=\mu^2,\,\, T_{--}=\mu^2$ will emerge  as the Virasoro constraints in the 
conformal gauge supplemented by the 
$t=\mu\tau$  condition fixing the residual conformal diffeomorphisms.

The Pohlmeyer reduction {(see,  e.g.,  ~\ci{GT} for an exposition of the general scheme)}
ammounts to using the $\bar G$ gauge freedom to
fix one component of $P_a$ 
\foot{This  construction of the reduced model 
is not unique in the case when the coset space $F/\bar G$
has rank  bigger than one, i.e. rank$(F)$ - rank$(G)$= 2, 3 , ... \ci{mirm}. Since 
 the case of our prime interest ($AdS_n \times S^n$) is based on rank one  cosets,
 here we shall discuss only this ``canonical'' choice. Let us note, however,
  that in order to apply the Pohlmeyer
 type reduction to the PCM with $G$ of rank $>1$ one also needs 
 to  fix values of other Casimirs besides $\tr(P_+P_+)$
and $\tr(P_-P_-)$. This more general reduction procedure \ci{mirm}
may  be useful in studying special 
solutions of such  models.}
\begin{equation}
P_+=\mu T\,,
\end{equation}
where $T$ is a particular   element of $\algp$, 
i.e. $T=(\rt,-\chi(\rt))$,  $\rt\in\algg$.
One can then parametrize $P_-$ as
\begin{equation}
P_-=\mu \bar g^{-1}T \bar g\,,  \la{ge}
\end{equation}
where $\bar g$ is a new field taking values in $\bar G$, 
i.e. having  the form $(g,\hat\chi(g))$.
 Then the original 
 $\bar G$ gauge symmetry  ($f \to k  f$,  $\ k \in  \bar G$) is broken
to the $\bar H$ gauge symmetry, where $\bar H\subset \bar G$ 
is a subgroup of elements preserving $T$:
the corresponding 
 subalgebra $\bar\algh$ is  the centralizer of $T$ in $\bar\algg$.

In addition, the introduction 
 of $\bar g$ brings in the new gauge symmetry:
   $\bar g$
and $\bar h \bar g$ with  $\bar h \in \bar H$  represent 
 the same $P_-$. The resulting (on-shell) formulation should thus have 
$\bar H\times \bar H$ gauge symmetry,\  $ \bar g \to \bar h \bar g \bar h'$. 

For a compact group  $G$  one can 
assume $\rt$ in $T=(\rt, -\chi(\rt))$  to be a nonvanishing element of 
the Cartan subalgebra of $\algg$. The centraliser
of $\rt$ in $\algg$ is the Cartan subalgebra $\algh$. The centralizer of $T$ in $\bar\algg$ is then
the same Cartan subalgebra embedded (twisted diagonally) into 
$\bar\algg\simeq \algg$. 

In addition to the field $\bar g$ in \rf{ge}
one finds also 
the 2d gauge field components $\bar A_+$ and $\bar A_-$ taking values in
 $\bar\algh$ and transforming under the  gauge groups (we shall assume the standard vector
  gauging here)  -- they emerge  from the other components of the current $f^{-1} d f$. 
After a partial gauge fixing the Pohlmeyer-reduced system for the 
PCM is then represented  by $\bar G/\bar H$ gauged WZW model with a
 potential.
For a  given   automorphism $\chi$, the  fields $\bar g= (g, \hat \chi (g)) $ 
and $\bar A_\pm= (A_\pm, 
 \chi ( A_\pm))   $ are uniquely 
determined by their first components; it is useful
to describe the reduced model in terms of  $g\in G $ and $A_\pm\in \algh$
(the action does not depend on $\chi$):
\bea
L_r=&-&\half\tr(g^{-1}\d_-g\ g^{-1}\d_+g)  + {\rm WZ\ term}  -\mu^2\tr(g^{-1}\rt g\rt)  \no \\ 
  &+& \tr \big( - A_+ \del_- g g^{-1}  +    A_-  g^{-1} \del_+ g 
   + g^{-1} A_+ g A_-   - A_+ A_- \big) \ .   \la{gw}
\eea
Here  
$A_\pm$ take values in the Cartan subalgebra $\algh\subset\algg$    and 
 $\rt\in \algg$ is a fixed element of $\algh$. The corresponding model
is the ``homogeneous   sine-Gordon'' model \ci{fernn}
that was studied   in the literature \ci{fern}.

The first nontrivial example 
is given by  $G=SO(3)$. In this case the reduced Lagrangian 
\eqref{gw} leads to  the complex sine-Gordon (CSG) model 
after eliminating the auxiliary fields $A_\pm$.
The CSG model \ci{pol} is known to be the  reduced theory for the coset 
$S^3=SO(4)/SO(3)$. The equivalence  is obvious if one uses the 
 representation $so(4)\simeq so(3)\oplus so(3)$.

Let us note that the PCM  for $G=SO(3)$ subject to the Virasoro constraints (i.e.
 the reduced model for strings on $S^3\times \fR^1$) also admits 
 an alternative  Fadeev-Reshetikhin 
 reduction~\cite{FR}.
  The FR theory is  formulated in terms of two unit 3-vectors or 4 independent variables
(related locally to the original current components)
and is  described   by   a
  first-order action. It thus  has  the same number of 
  degrees of freedom (two in  a second-derivative form) 
  as in  the CSG model.
  However, in  contrast  to  CSG,  the FR model is not explicitly 
    2d Lorentz invariant. 
   The  
     CSG and the FR models  which are both related to the same PCM  equations of motion
     with the Virasoro constraints imposed 
    should  then be related by a (nonlocal)
field redefinition.\foot{It might   be possible to  consider the FR and CSG models as originating from 
    two different 
   gauges of the $G\times G  \ov G$ coset sigma model.}

In the next section  we shall   consider the reduced model
 for the superstring on $AdS_3\times S^3$.
The bosonic part of the $AdS_3\times S^3$ superstring  
sigma     model~\cite{adst} is the direct sum of the coset models
of the type~\eqref{homs}, i.e.  
$AdS_3\times S^3\simeq SU(1,1)\times SU(2)$ 
can be  represented as a coset~\eqref{homs} with $G=SU(1,1)\times SU(2)$.

\section{Superstring theory  on $AdS_3\times S^3$}\label{sec:3}

The Green-Schwarz superstring on $AdS_3\times S^3$
supported by   RR 3-form flux  can be  formulated as 
a coset model for the supercoset~\cite{Ram-Raj,adst}
\begin{equation}
\label{coset-main}
\frac{PSU(1,1|2)\times PSU(1,1|2)}{SU(2)\times SU(1,1)}
\end{equation}
The  superalgebra $psu(1,1|2)$ of  $PSU(1,1|2)$ is 
represented by $(2|2)\times (2|2)$ traceless supermatrices
satisfying an appropriate reality condition; the  quotient is over the central subalgebra
 generated by the unit matrix (for details see,  e.g.,~\cite{GT}). 
 This algebra (as well as $psu(2,2|4)$ and its higher-dimensional analogs)
admits a $Z_4$-grading \cite{BBHZZ}. This grading appears to be 
 extremely useful in studying such  sigma-models and their
  Pohlmeyer-type reductions. In particular, the formulation of superstrings on
$AdS_2\times S^2$ or $AdS_5\times S^5$ is most convenient in terms of $Z_4$-decomposition of
the algebra-valued currents.

\subsection{$Z_4$ grading of the superalgebra} 

In the present case  we need a $Z_4$ decomposition of the superalgebra 
 $\alghf=psu(1,1|2)\oplus psu(1,1|2)$. The grading we are interested in is 
different from the one induced by the standard 
grading on each term in the sum:  the one
 we are looking for mixes the two terms.

To identify the required grading in terms of matrix representation let us consider first
the bosonic part given by a direct sum of two copies of $su(1,1) \oplus su(2)$.
 The degree zero component is formed by elements of the form $(a,-a^t)$ 
 with $a\in su(1,1)\oplus  su(2)$ while the degree $2$ component is formed
by  $(a,a^t)$. These two components are  orthogonal to each other and satisfy
\begin{equation}
\commut{\algf_0}{\algf_0}\subset \algf_0\,,\qquad \commut{\algf_0}{\algf_2}\subset \algf_2\,, \qquad
\commut{\algf_2}{\algf_2}\subset \algf_0\,,
\end{equation}
so that they can be identified with the even-degree components of the $Z_4$-decomposition.
Moreover, the degree zero component is obviously isomorphic to $su(1,1)\oplus su(2)$,  i.e. to 
the denominator of the coset~\eqref{coset-main}.

To extend the grading to the fermionic components it is useful to consider first the grading
of the complexified algebra $\alghf^\fC=psl_\fC(2|2)\oplus psl_\fC(2|2)$ and to represent its elements 
by $8\times 8$ block-diagonal matrices of the form
\begin{equation}
 \left(\begin{array}{cccc}
 a &\alpha &0& 0\\
\beta &b &0& 0\\
0 &0& c &\gamma \\
0 &0& \delta & d
\end{array}
\right) \ . 
\end{equation}
Here $a,c,b,d$ are $2\times 2$ bosonic matrices from $sl(2)$;
$\alpha,\beta,\gamma,\delta$ are complex fermionic matrices.
The antiautomorphism determining the $Z_4$ structure is given by
\begin{equation}
M^\Omega=-\mathbf{K}^{-1}M^{st}\mathbf{K}\,,
\qquad \mathbf{K}=
\left(
\begin{array}{cc}
0&K\\
K&0
\end{array}
\right)\,,
\qquad 
K=\left(
\begin{array}{cc}
\id&0\\
0&\id
\end{array}
\right)\,,
\end{equation}
where $\id$ is the unit $2\times 2$ matrix and $~^{st}$ denotes the transposition of the supermatrices.
More explicitly,  one has
\begin{equation}
 \left(\begin{array}{cccc}
 a &\alpha &0& 0\\
\beta &b &0& 0\\
0 &0& c &\gamma \\
0 &0& \delta & d
\end{array}
\right)^\Omega=
-\left(\begin{array}{cccc}
 c^t &-\delta^t &0& 0\\
\gamma^t &d^t &0& 0\\
0 &0& a^t &-\beta^t \\
0 &0& \alpha^t & b^t
\end{array}
\right)
\end{equation}
The $Z_4$ components $\alghf^\fC_k$ are then identified as the eigenspaces of $\Omega$, i.e. 
$M^\Omega= \imath^k M$ for $M\in \alghf^\fC_k$ so that 
$\alghf^\fC=\alghf^\fC_0\oplus \alghf^\fC_1\oplus\alghf^\fC_2\oplus\alghf^\fC_3$.

To obtain $psu(1,1|2)\oplus psu(1,1|2)$ one needs to impose the reality condition $M^*=-M$ where
$^*$ is an antilinear antiautomorphism defined as 
\begin{equation}
\left(
\begin{array}{cc}
a &\alpha\\
\beta & b
\end{array}
\right)^*=
\left(
\begin{array}{cc}
\Sigma a^\dagger \Sigma & -\imath \Sigma \beta^\dagger\\
-\imath \alpha^\dagger \Sigma & b^\dagger
\end{array}
\right)\,,\qquad\quad
\Sigma=\left(
\begin{array}{cc}
1&0\\
0&-1
\end{array}
\right)\,,
\end{equation}
and analogously for the second copy of $psl(2|2)$. Here $^\dagger$ 
denotes the ordinary hermitean conjugation.\foot{Note that one can also take $\Sigma$ to be a unit matrix. This would correspond to 
describing strings on $S^3\times S^3$ with the signature $(3,3)$.}
 In terms of the components
the reality condition reads as
\begin{equation}
\Sigma a^\dagger \Sigma=-a\,,\qquad b^\dagger=-b\,,\qquad 
\imath \Sigma \beta^\dagger=\alpha\,, \qquad \imath \Sigma \delta^\dagger=\gamma\,,
\end{equation}
along with the same conditions for the components of the second copy of $sl(2|2)$ (i.e. for $c,d,\gamma,\delta$).

It turns out that the $Z_4$ decomposition of $psl_\fC(2|2)\oplus psl_\fC(2|2)$ is compatible
 with the above  reality condition in the sense that if $M\in \alghf^\fC_k$,
  i.e. $M^\Omega=\imath^k  M$ then 
$M^*\in \alghf^\fC_k$. This implies that $\Omega$ induces the $Z_4$ decomposition of 
$\alghf=psu(1,1|2)\oplus psu(1,1|2)$ 
\begin{equation}
\label{decompos}
\alghf=\alghf_0\oplus \alghf_1\oplus\alghf_2\oplus\alghf_3\,,
\qquad  \qquad \commut{\alghf_i}{\alghf_j}\subset \alghf_{i+j\,\mathrm{mod}\,4}\ . 
\end{equation}
The subspace $\alghf_k$ is given by the intersection of $\alghf\subset\alghf^\fC$ and
$\alghf^\fC_k\subset \alghf^\fC$.\foot{Let us note that $\Omega$ is not an antiautomorphism
of $\alghf$ as it maps elements satisfying the reality condition to those which do not.
That is why it  is useful to consider $\alghf^\fC$ in order to identify the grading on $\alghf$.}


Once the $Z_4$-grading is identified,
 the construction of the superstring sigma model
coincides with that for  $AdS_5\times S^5$ and $AdS_2\times S^2$ cases in  \ci{BBHZZ}. 
The Lagrangian 
 is written  in terms of the $Z_4$-components of the current $J_\pm=\hat f^{-1}\d_\pm 
 \hat f$
\begin{equation}
 J_\pm =\A_\pm+P_\pm+Q_{1\pm} +Q_{2\pm}\,,\qquad \A \in \alghf_0,\quad Q_1\in \alghf_1,
\quad P\in\alghf_2,\quad Q_2\in \alghf_3\ .
\end{equation}
Explicitly, in the conformal gauge  ($\mathrm{STr}$ is the supertrace) 
\begin{equation}\la{lak}
 L_{\rm GS}= \mathrm{STr}\big[P_+ P_- + \half (Q_{1+}Q_{2-}-Q_{1-}Q_{2+}) \big]\,.
\end{equation}
The Virasoro constraints  are  $\mathrm{STr}(P_+P_+) =0$ and
$\mathrm{STr}(P_-P_-)=0 $.  The   GS action  (before conformal gauge fixing) 
is invariant under the fermionic
$\kappa$-symmetry. This invariance can be  partially fixed by the following gauge condition \ci{GT}:
\begin{equation}\la{kg} 
 Q_{1-}=0\,,\qquad \qquad Q_{2+}=0\,.
\end{equation}

\subsection{Pohlmeyer reduction}

Given  a superstring action  written 
 in terms of the $Z_4$ components of the currents,  all 
the remaining steps of the Pohlmeyer reduction 
 are the same  as in  
the $AdS_5\times S^5$ or $AdS_2\times S^2$ cases discussed in \cite{GT}.
Here we give a short review of the procedure concentrating on 
the subtleties of the $AdS_3\times S^3$ case.

The Pohlmeyer reduction is performed in terms of the $Z_4$-components
$\A,Q_1,P,Q_2$ of the current 
$J_\pm=f^{-1}\d_\pm f$.  The components $Q_{1-}$ and $Q_{2+}$ are set to zero 
as partial $\kappa$-symmetry gauge fixing. Using the $\algf_0$-gauge symmetry one can
assume that $P_+=p_+T$ where $p_+=p_+(\sigma)$
is some scalar function and
$T$ is a fixed element of $\alghf_2$. 
In the case at hand there are inequivalent choices of $T$. One can,  for instance,  take 
$T=(\rt,-\rt^t)$ with $\rt|_{su(2)}=0$ or, alternatively, 
 $\rt|_{su(1,1)}=0$. These choices are clearly inequivalent.
The ``nondegenerate'' choice we are going to utilize 
is the one where both $su(1,1)$ and $su(2,2)$ parts are nonvanishing.
Namely, we take (cf.  \cite{GT})
\begin{equation}\la{lq}
 T=diag(\rt,\rt^t)\,,\qquad \rt=\frac{\imath}{2}diag(1,-1,1,-1)\,.
\end{equation}
Note that in this matrix representation $\rt$  coincides with $T$ used 
 in the $AdS_2\times S^2$ case in \cite{GT}.

The choice of  $T$ in \rf{lq}  induces the decomposition
 $\alghf=\alghf^\oo\oplus \alghf^\pp$ 
 in each of the two  $psu(1,1|2)$ sectors. More precisely,
\begin{gather}
 \alghf=\alghf^\pp\oplus\alghf^\oo \ , \qquad 
  P^\pp\zeta^\pp=\zeta^\pp\, ,   \qquad  P^\pp{\chi^\oo}=0\ , \\
\la{mkm}
  \zeta^\pp\in \alghf^\pp \ , \qquad  \chi^\oo\in {\alghf}^\oo\ , \qquad
  P^\pp\equiv -\commut{T}{\commut{T}{\cdot\ }}\  . 
\end{gather}
Note  that $\alghf^\pp=\im(ad(T))$ and $\alghf^\oo=\ker(ad(T))$; moreover, 
 for $\zeta^\pp\in \alghf^\pp$ one also has
 $\scommut{T}{\zeta^\pp}=0$.

In each $psu(1,1|2)$ sector the decomposition $\alghf=\alghf^\oo\oplus \alghf^\pp$ is identical
to that in the $AdS_2\times S^2$ case in \cite{GT}.
However, the choice of the subspaces
\begin{equation}
 \alghf_0^\oo=\bar\algh\,,\quad \alghf_0^\pp=\bar\algm\,,\quad \alghf_{1,3}^\pp\,,\quad  \alghf_{1,3}^\oo
\end{equation}
here is  different from \ci{GT} as $Z_4$-grading is defined in a different way and mixes the two $psu(1,1|2)$
sectors.  In particular, the subalgebra
$\bar\algh=\alghf_0$ is two dimensional, $\bar\algh\simeq u(1)\oplus u(1)$,  and a useful choice of its  basis is
\begin{equation}
 h^{(A)}=diag(\imath,-\imath,0,0,-\imath,\imath,0,0)\,,
\qquad 
 h^{(S)}=diag(0,0,\imath,-\imath,0,0,-\imath,\imath,)\,.
\end{equation}
Next, one uses the Virasoro constraint $\str{(P_+P_+)}=0$ and the residual 
conformal invariance 
to set $p_+=\mu$ for some constant $\mu$ so that $P_+=\mu T$.
Introducing the $\bar G$-valued field $\bar g$ ($\bar G\subset \hat F$  is a
 subgroup corresponding to the  subalgebra $\alghf_0\simeq  su(1,1)\oplus su(2)$, 
 i.e. $\bar G\equiv SU(1,1)\times SU(2)$) one solves the equation of motion 
 $\d_+ P_- +\commut{ \A_+}{P_-}=0$ and the Virasoro constraint
$\str(P_-P_-)=0$ by
\begin{equation}
 P_-=\mu \bar g^{-1}T\bar g\,,
\end{equation}
where one  again used the remaining conformal transformation freedom.
Finally, solving the remaining equations of motion
by choosing 
\begin{equation}
\A_-=\bar A_-\,,\qquad \A_+=\bar g^{-1}\d_+\bar g+\bar g^{-1} \bar A_+\bar g
\end{equation}
one ends up with only the Maurer-Cartan equation imposed 
on the $\alghf$-connection $J$
parametrised in terms of the new fields:
 $\bar G$-valued field $\bar g$,  the $\bar\algh=\alghf_0^\oo$-valued
fields $\bar A_+, \bar A_-$ and the fermionic fields $Q_1,Q_2$. In this parametrization the Maurer-Cartan equation
is invariant under
the $\bar H\times \bar H$  local symmetry 
(recall that in our case $\bar H\simeq U(1)\times U(1)$ is 
the Lie group whose Lie algebra is $\bar\algh=\alghf_0^\oo$). 


Finally,  one uses the residual kappa-invariance to set to zero the components
$Q_{1+}^\oo$ and $(\bar g Q_{2-}\bar g^{-1})^\oo$. The remaining components of the fermionic currents
 are parametrized in terms of the new fermionic fields $\Psr,\Psl$ taking values in $\alghf_1$ 
 and $\alghf_3$  respectively:
\begin{equation}\la{psis}
 Q_{1+}^\pp=\sqrt{\km}\,\Psr\,,\qquad 
 (\bar g Q_{2-} \bar g^{-1})^\pp=\sqrt{\km}\,\Psl\,.
\end{equation}
Using $\bar H\times \bar H$ local symmetry one can satisfy the  following constraints:
\begin{equation}
\begin{aligned}
 \label{A-const}
\tau(\bar A_+)&=(\bar g^{-1}\d_+\bar g+\bar g^{-1}\bar A_+\bar g)_\algh 
-\half\commut{\commut{T}{\Psr}}{\Psr}\,,
 \\
\bar A_-&= ( \bar g\d_-\bar g^{-1}+\bar g\,\tau({\bar A}_-)\bar g^{-1})_\algh 
-\half\commut{\commut{T}{\Psl}}{\Psl}\,.
\end{aligned}
\end{equation}
where $\tau$ is an automorphism of $\bar\algh$ which is assumed to preserve the inner 
product (i.e. the trace).
This automorphism is introduced for generality 
to make  the resulting theory  having  a nonsingular expansion
around the natural vacuum $\bar g=\id$. This can be achieved 
by  choosing  $\tau (\bar A)=-\bar A$ which is an automorphism of
$u(1)\oplus u(1)$  (this will correspond to axial instead of vector gauging).
  The residual gauge transformations (i.e. the transformations preserving the
 Maurer-Cartan equations and the constraints \eqref{A-const}) read as:
\begin{gather}
\label{gs+f}
\bar g \to h^{-1}\bar g\,\, \hat\tau(h)\,, \quad \bar A_{+}\to h^{-1}\bar A_+h +h^{-1}\d_+ h\,, \quad   \bar A_{-}\to 
{h}^{-1}\bar A_{-}{h}+{h}^{-1}\d_-  h\,,\\
\label{Q-gs}
\Psr\to {\hat\tau(h)}^{-1} \Psr\,\, \hat\tau(h) \,, \qquad \qquad \Psl\to h^{-1} \Psl h\,. 
\end{gather}
The Maurer-Cartan equations and the constraints~\eqref{A-const} can  then be obtained 
 from the following  local Lagrangian:\foot{{The equations similar to
those contained in the Maurer-Cartan equations 
appeared in a different context in \cite{Aratyn} and are 
 formally invariant under a 2d
 supersymmetry. However, besides  these equations the Lagrangian \eqref{L-tot}
  leads  also to the constraints \eqref{A-const} that are not,  in general, 
   invariant under the
  supersymmetry transformations (cf. \cite{GT}).}}
\begin{multline}
\label{L-tot}
L_{tot}=L_{\rm gWZW}+\mu^2\,\mathrm{STr}(\bar g^{-1}T \bar g T )
+\mathrm{STr}\left(\Psl {T} \bar D_+\Psl+
\Psr {T} \bar D_-^\tau\Psr\right)+\mu\,\mathrm{STr}\left(
\bar g^{-1}\Psl \bar g\Psr\right),
\end{multline}
where
\begin{equation}
\bar D_+\Psl=\d_+\Psl+\commut{\bar A_+}{\Psl}\,,\qquad 
\bar D_-^\tau\Psr=\d_-\Psr+\commut{\tau(\bar A_-)}{\Psr}\,,
\end{equation}
and  $\Psr,\Psl $ are constrained by the condition that they anticommute with $T$
(i.e. take values in $\alghf_{1,3}^\pp$). 
$L_{gWZW}$ which depends only on the bosonic fields  is given explicitly by
\begin{multline} \la{gaui}
L_{\rm gWZW}  = \half\str(\bar g^{-1}\d_+\bar g \bar g^{-1}\d_-\bar g) +\text{WZ-term}\\
 +\str(  \bar A_+\,
 \d_- \bar g \bar g\inv -
 \tau(\bar A_-) \,\bar g\inv\del_+ \bar g   - \bar g\inv \bar A_+ \bar g \, \tau(\bar A_-)  + \bar A_+ \bar A_- 
 \big)\,.
\end{multline}
Here  the supertrace in the bosonic terms 
accounts for the relative minus sign  in the contributions of the 
${S^3}$ and ${AdS_3}$ parts  (leading to the correct final signs).


\bigskip

\subsection{Reduced Lagrangian in terms of independent  degrees of freedom}

Similarly to the purely bosonic case,
 the Lagrangian \rf{L-tot}  of the reduced model 
 can be usefully parametrized in terms of the bosonic
  fields taking values in one copy of $psu(1,1|2)$ only.
   Namely,  let $g$ be an  $SU(1,1)\times SU(2)$-valued field, 
   $A_\pm$ the $u(1)\oplus u(1)$-valued
gauge fields, and $\Psr^\prime,\Psl^\prime$ take values in the fermionic 
part of the ``parallel'' subspace  of the first $psu(1,1|2)$, i.e.  
\begin{gather}
 g=\left(
\begin{array}{cc}
 g_A &0\\
0 &g_S
\end{array}
\right)
\,,
\qquad
A_\pm=
\left(
\begin{array}{cc}
 A^A_\pm &0\\
0 &A^S_\pm
\end{array}
\right)
\,,
\qquad
\Psi^\prime_{1,2}=
\left(
\begin{array}{cc}
 0& \psi_{1,2}\\
\imath \psi_{1,2}^\dagger \Sigma &0\\
\end{array}
\right)\,.
\label{Psipsi}
\end{gather}
Here $A$ and $S$  refer to the $AdS$  and the sphere parts, i.e.
 $g_A$ and $g_S$ are in the fundamental representations of 
$SU(1,1)$ and $SU(2)$ respectively,
$A^A_\pm=a^A_\pm \,diag(\imath \Sigma,0)$, $A^S_\pm=a^S_\pm \,diag(0,\imath \Sigma)$ and $\psr,\psl$
are antidiagonal complex fermionic matrices. Recall that $\Sigma=diag(1,-1)$ and
$\rt=\frac{\imath}{2}\,diag(\Sigma,\Sigma)$. 

More explicitly, let us choose the following basis in $su(1,1)$ and $su(2)$ 
in terms of the Pauli matrices:
$\bar R_1= \s_1, \ \bar R_2= i \s_3, \ \bar R_3=\s_2,$  and 
$ R_1= i \s_1, \  R_2= i \s_3, \  R_3=i\s_2$ (see Appendix~\bref{app} for details).
{To simplify the presentation let us first consider the case of $\tau=\id$}.
One can  parametrize 
 the group valued
  field $g$ in terms of  the Euler angles $\phi,\chi$ and $\varphi,\theta$ as
\begin{equation}\label{param}
  g_A= \exp{(\ha \chi\bar R_2)}\exp{(\phi \bar R_1)}\exp{(\ha\chi\bar R_2)},\quad
  g_S=\exp{(\ha\theta R_2)}\exp{(\varphi  R_1)}\exp{(\ha\theta R_2)}\,.
\end{equation} 
Explicitly, 
\begin{equation}
g_A=
\left(\begin{array}{cc}
 e^{\imath\chi}\cosh \phi &\sinh \phi\\
\sinh \phi & e^{-\imath\chi}\cosh \phi
\end{array}
\right)\,,\qquad
g_S=
\left(\begin{array}{cc}
 e^{\imath \theta}\cos \varphi &\imath \sin \varphi\\
\imath \sin \varphi & e^{-\imath \theta}\cos \varphi
\end{array}
\right)\,.
\end{equation}
One  can then solve for the gauge fields  using their equations 
\bea
\label{A-const-v}
&& A_+=(\hat A_+)_\algh\,,  \qquad \hat A_+ \equiv  g^{-1}\d_+g+g^{-1}A_+g 
-\half\commut{\commut{t}{\Psr^\prime}}{\Psr^\prime}\,,
 \\
 \label{hatA-v}
&&  A_-=(\hat A_-)_\algh \,, \qquad
\hat A_- \equiv  g\d_-g^{-1}+gA_-g^{-1} 
-\half\commut{\commut{t}{\Psl^\prime}}{\Psl^\prime}\,,
\eea
following from the Lagrangian \eqref{L-tot} with $\tau=\id$.
The fermionic terms entering the constraints give
\begin{equation}
\label{ferm-term}
\half\commut{\commut{t}{\Psi_1^\prime}}{\Psi_1^\prime}=(\alpha\beta-\gamma\delta)(\bar R_2-R_2)\,,
\qquad
\half\commut{\commut{t}{\Psi_2^\prime}}{\Psi_2^\prime}=(\l\nu-\rho\sigma)(\bar R_2-R_2)\,,
\end{equation}
where we have introduced the real components of the fermions in  \eqref{Psipsi} as 
\begin{equation}
\label{ferm-param}
\psr=\left(\begin{array}{cc}
 0 &\alpha+\imath \beta\\
\gamma+\imath\delta& 0	
\end{array}
\right)\ ,  \qquad
\psl=\left(\begin{array}{cc}
 0 &\l+\imath \nu\\
\rho+\imath\sigma& 0	
\end{array}
\right)\,.
\end{equation}
One then finds
\begin{equation}
A^A_+=\frac{\d_+\chi(1+\cosh2\phi)-2(\alpha\beta-\gamma\delta)}{2(1-\cosh{2\phi})}\bar R_2\,, \quad
A^S_+=\frac{\d_+\theta(1+\cos2\varphi)+2(\alpha\beta-\gamma\delta)}{2(1-\cos{2\vp})}R_2\,,
\end{equation}
and similar expressions  for $A_-$
with  $\d_+\chi \to -\d_-\chi$ and $\alpha\beta-\gamma\delta   \to    \l \nu - \rho \sigma $.

Using the equations of motion for $A_\pm$ one can write the reduced Lagrangian in the form
\bea
&&L_{tot}=
\half\tr(g_A^{-1}\d_+g_A g_A^{-1}\d_-g_A)-
\half\tr(g_S^{-1}\d_+g_S g_S^{-1}\d_-g_S) + \text{potential}
\no\\
&&+\text{fermionic kinetic term }
+ \text{fermionic interaction term} \la{hi}\\
&&+\tr\big(A^A_+ [g_A,\psi_1](\d_-g_A g_A^{-1}+\half\commut{\commut{\rt}{\Psi^\prime_2}}{\Psi^\prime_2})_{AdS}
-A^S_+[g_S,\psi_1](\d_-g_S g_S^{-1}+\half\commut{\commut{\rt}{\Psi^\prime_2}}{\Psi^\prime_2})_{S}\big)\,.
\nonumber
\eea
The bosonic part of  the  Lagrangian that comes from the WZW  and potential
 terms (i.e. terms 
not involving $A_\pm$) is 
\bea 
&&L_1= 
   \dpp \varphi \dmm \varphi  +   \ha (1 + \cos 2  \varphi ) \ \dpp \theta \dmm \theta \no \\
&&   +\
   \dpp \phi \dmm \phi       - \ha ( 1 + \cosh 2   \phi ) \ \dpp \chi  \dmm \chi
   + \frac{\m^2}{2} (\cos 2\varphi  - \cosh 2\phi) \ . 
\eea
The  fermionic interaction term is found to be
$\str(g^{-1}\Psl^\prime g\Psr^\prime)
=
-2 {\rm Im}[\tr(g_A^{-1}\psi_2 g_S \psi_1^\dagger\Sigma)]$
and  together with the fermionic kinetic terms they give 
\bea
&&L_{2} = 
\alpha\d_-\alpha+\beta\d_-\beta+\gamma\d_-\gamma+\delta\d_-\delta+\l\d_+\l+\nu\d_+\nu
+\rho\d_+\rho+\sigma\d_+\sigma\no  \\
&&-2\mu\Big(
\sinh{\phi}\sin{\varphi}(\l \gamma+\nu\delta-\rho\alpha-\sigma\beta) 
+\cosh{\phi}\cos{\varphi} \big[ \cos{(\chi+\theta)}(\rho\delta-\sigma\gamma\no\\
&& -\l\beta+\nu\alpha) -\sin{(\chi+\theta)}(\l\alpha+\nu\beta+\rho\gamma+\sigma\delta)\big]\Big) \,.
\eea
Finally, the terms that originate from the elimination of $A_\pm$ (third line of \rf{hi}) 
are 
\bea
&&L_{3} = 
  \frac{[\d_+\chi\ (1+\cosh 2\phi)\ -2 (\alpha\beta-\gamma\delta)]
 [\d_-\chi\ (1+\cosh 2\phi)\ +2(\l\nu-\rho\sigma)]}{2(\cosh 2\phi -1) } \no  \\
&&+\  \frac{[\d_+\theta\ (1+\cos 2\varphi)\ +2(\alpha\beta-\gamma\delta)][\d_-
 \theta\ (1+\cos 2\varphi)\ -2(\l\nu-\rho\sigma)]}{2(1-\cos 2\varphi) }\ . 
\eea
Then the Lagrangian \rf{hi} becomes 
\begin{equation}\label{L-tot-short}
L_{tot}= L_1 + L_2 + L_3 \equiv L_B + L_F \ . 
\end{equation} 
The purely bosonic terms in $L_1 $ and $L_3$ combine into the direct sum 
of the CSG   action and its ``hyperbolic'' counterpart
which is the reduced Lagrangian for the bosonic string in  $AdS_3 \times S^3$:
\begin{equation}
 \label{hgi}
   L_B =
   \dpp \varphi \dmm \varphi  +  \cot^2 { \varphi }\ \dpp \theta \dmm \theta
   +
   \dpp \phi \dmm \phi  +   \coth^2 { \phi }\ \dpp \chi  \dmm \chi
    +\frac{\m^2}{2} (\cos 2\varphi  - \cosh 2\phi) \,,
\end{equation}
while the fermionic ones give:
\begin{multline}\la{lll}
L_{F}= 
L_2-\cot^2\varphi\ [\d_+\theta(\lambda\nu-\rho\sigma)-\d_-\theta(\alpha\beta-\gamma\delta)]
+\coth^2\phi\ [\d_+\chi(\lambda\nu-\rho\sigma)-\d_-\chi(\alpha\beta-\gamma\delta)]\\
-(\alpha\beta-\gamma\delta)(\lambda\nu-\rho\sigma)[\frac{1}{\sin^2\varphi}+\frac{1}{\sinh^2\phi}]\,.
\end{multline}
For the Lagrangian $L_{tot}$ the point $\varphi=\phi=0$ 
which is a minimum of the potential  is a  singular point of the kinetic-term.\foot{This  
point is still a regular expansion point for the corresponding Hamiltonian, 
assuming the momenta  of $\theta$ and $\chi$ are constant in the vacuum.}
At the same time   the regular point of the kinetic term 
$\varphi=\pi/2,\,\,\phi=\imath\pi/2$ is a maximum of the potential. 
One can  by-pass this  complication as in the 
 purely bosonic case -- by 
using the axial gauged gWZW theory instead of the  vector gauged one.


To find  the axial gauging analog of the above reduced Lagrangian   \rf{L-tot-short}
we are to  take  $\tau(a)=-a,\,a\in\algh$ in \eqref{L-tot}.
Using this  asymmetric gauge also affects the parametrization of 
the group element: now one is to use 
\begin{equation}
 g=\hat\tau(g_2)g_1g_2\,,
\end{equation}
leading to (cf. \eqref{param})
\begin{equation}
 g_A= \exp{(-\ha \chi\bar R_2)}\exp{(\phi \bar R_1)}\exp{(\ha\chi\bar R_2)},\quad
  g_S=\exp{(-\ha\theta R_2)}\exp{(\varphi  R_1)}\exp{(\ha\theta R_2)}\,.
\end{equation}
One can then redo  the same steps as  above and get the corresponding 
 Lagrangian in terms of the physical degrees of freedom only.
  Details of this are given in the Appendix~\bref{app}. 
  It turns out that similarly to the purely bosonic CSG case 
   the resulting Lagrangian can be obtained directly 
  from the vector-gauged $L_{tot}$ by an appropriate ``analytic continuation''. Namely,  
  transforming the variables according to
\begin{equation}
 \varphi\to \varphi+\frac{\pi}{2}\,, \quad \phi\to \phi+\imath\frac{\pi}{2}\,, \qquad
\theta\to -\theta\,,\quad \chi \to -\chi\,, 
\end{equation}
and redefining the coupling as $\mu\to -\imath\mu$
one gets the resulting ``dual'' Lagrangian
\begin{multline}
\label{L-tau}
   L_{tot}^ {axial} =
  \dpp \varphi \dmm \varphi  +  \tan^2 { \varphi }\ \dpp \theta \dmm \theta
  +
  \dpp \phi \dmm \phi  +   \tanh^2 { \phi }\ \dpp \chi  \dmm \chi
   +\frac{\m^2}{2} (\cos 2\varphi  - \cosh 2\phi)\\
+\alpha\d_-\alpha+\beta\d_-\beta+\gamma\d_-\gamma+\delta\d_-\delta+\l\d_+\l+\nu\d_+\nu
+\rho\d_+\rho+\sigma\d_+\sigma\\
+\tan^2 \varphi\  [\d_+\theta(\lambda\nu-\rho\sigma)-\d_-\theta(\alpha\beta-\gamma\delta)]
-\tanh^2 \phi\ [\d_+\chi(\lambda\nu-\rho\sigma)-\d_-\chi(\alpha\beta-\gamma\delta)]\\
-(\alpha\beta-\gamma\delta)(\lambda\nu-\rho\sigma)[\frac{1}{\cos^2\varphi}-\frac{1}{\cosh^2\phi}]
-2\mu\Big(
\cosh{\phi}\cos{\varphi}(\l \gamma+\nu\delta-\rho\alpha-\sigma\beta) \\
+\cosh{\phi}\cos{\varphi} \big[ \cos{(\chi+\theta)}(-\rho\delta+\sigma\gamma
 +\l\beta-\nu\alpha) -\sin{(\chi+\theta)}(\l\alpha+\nu\beta+\rho\gamma+\sigma\delta)\big]\Big) \,.
\end{multline}
Note that in order to obtain this Lagrangian directly from \eqref{L-tot} with $\tau(a)=-a$
one also needs to redefine the fermions as follows:
 $\alpha\to-\delta,\delta\to\alpha,\beta\to\gamma,\gamma\to-\beta$.

\bigskip  

Since (as follows from \rf{lll} and \rf{L-tau}) 
we may identify  the fermions     $\alpha,\beta,\gamma,\delta$ and $\l,\nu,\rho,\sigma$
with  2d Majorana-Weyl spinors,  a   natural question then is if the total 
reduced  Lagrangian  has a 2d supersymmetry, i.e. if  it 
can  be interpreted as a supersymmetric extension  of \rf{hgi}.
This is indeed possible  for  a consistent truncation 
of $L_{tot}$ (for definiteness let us consider  \eqref{L-tot-short})  found by setting 
$\chi=\theta=0,\  \ \lambda=\gamma=\sigma=\beta=0$  
which produces the reduced Lagrangian 
for the $AdS_2 \times S^2$  superstring 
\ci{GT}:
\begin{multline}\la{taa}
L_{trunc.}=\d_+\vp\d_-\vp+\d_+\phi\d_-\phi  + {\m^2\ov 2 }(\cos 2\vp - \cosh 2\phi )
+  \alpha\d_-\alpha  +\delta\d_-\delta \\
+\nu\d_+\nu
+ \rho\d_+\rho          
- 2\mu\big[
\cosh{\phi}\ \cos{\varphi}\ (\nu\alpha+\rho\delta )+
\sinh{\phi}\ \sin{\varphi}\ (\nu\delta-\rho\alpha) 
\big] \  . 
\end{multline}
This Lagrangian is  equivalent \ci{GT}   to the $N=2$ supersymmetric sine-Gordon   Lagrangian\ci{susy}:
 \begin{multline}
L= \d_+ \Phi \d_- \Phi^*  - | W'(\Phi)|^2  
+\psi_{_L}^* \d_+ \psi_{_L} 
+\psi_{_R}^* \d_-  \psi_{_R}
+ \big[
 W''(\Phi)\psi_{_L}\psi_{_R} +   W^*{}''(\Phi^*) \psi^*_{_L}\psi_{_R}^*\big]\ , 
\la{comp} 
 \end{multline}
where 
$$\Phi= \vp + i \phi\ ,  \quad 
\psi_{_L}=\nu -\imath \rho\ ,\ \quad \psi_{_R}=-\alpha+\imath\delta\ ,  \quad 
 W= {\m  }  \cos \Phi \ . $$
At the same time,  both  the CSG model and  its ``hyperbolic''  analog 
admit $N=2$ supersymmetric extensions \ci{naps}
based on interpreting  $\xi\equiv \ln \cos \vp+i\theta$   and
 $\eta\equiv \ln \cosh \phi +i \chi$ as
complex scalar  components of chiral superfields   and using that 
$$d\vp^2 + \cot^2\vp d\theta^2= { \del^2 K \ov \del \xi \del \bar \xi} d \xi d \bar \xi\ , \ \ \ \ \ \  
d\phi^2 + \coth^2\phi d\chi^2= { \del^2 K' \ov \del \eta \del \bar \eta} d \eta d \bar \eta\ . $$
Then $K$ and $K'$ are the corresponding Kahler potentials,   while 
the two  superpotentials  are $ \mu e^\xi$  and $\mu e^\eta$. The resulting $N=2$ supersymmetric 
Lagrangian  is,  however, a direct sum of the two decoupled $N=2$  theories 
and thus cannot be equivalent to the above $L_{tot}$ (in particular, it does not admit
the above  $N=2$ SG truncation \rf{taa}). 

To show that $L_{tot}$ \rf{L-tot-short} or  \rf{L-tau} 
 has $N=2$  supersymmetry   one may try to  use 
non-standard  types of $N=2$ superfields (see, e.g.,  \ci{roc}).
While the sigma-model part of \rf{hgi}  admits straightforward $N=1$ 
supersymmetrization, incorporating the potential terms appears to  be non-trivial
(cf. \ci{papd} and refs. there).  The  existence of  2d supersymmetry of the reduced Lagrangian 
$L_{tot}$ thus remains an open   problem.

\section{Comments  on Pohlmeyer reduction of strings on $AdS_ n\times S^n$}\label{sec:4}

In this section   we shall make few general comments clarifying 
 some aspects 
of Pohlmeyer reduction of strings on $AdS_ n\times S^n$  spaces  and 
extending the discussion in \ci{GT}.

\subsection{Relation to Pohlmeyer reduction in  the pure $AdS_ n$  case}

Considering strings moving on $AdS_ n\times S^n$  we have assumed that the 
conformal gauge (Virasoro) condition $T^{AdS}_{\pm \pm} + T^{S}_{\pm \pm}=0$
is satisfied by $T^{S}_{\pm \pm}=\mu^2$, \ $T^{AdS}_{\pm \pm}= - \mu^2$.
Indeed,  if strings  move on  a sphere  their stress tensor must be positive 
and by residual conformal transformation can be made constant. However, 
there is a  special subclass of strings  which are localised 
on the sphere   and  move only in $AdS_n$; then 
we should have    $T^{S}_{\pm \pm}=0$, \ $T^{AdS}_{\pm \pm}= 0$.
In the context of string theory in $AdS_ n\times S^n$  this 
special  case should be viewed  as a limit $\mu\to 0$ 
of the general case.\foot{In the  case of $AdS_ n\times S^n$ 
the standard and natural choice of the expansion point or vacuum is the BMN one, i.e. 
the geodesic $t= \mu \tau, \ \ \psi= \mu \tau$,  implying a non-zero value for  $\mu$.}
Still, since in the non-compact $AdS_n$ case the condition 
$T^{AdS}_{\pm \pm}= 0$ has,  in general,   nontrivial solutions, 
   one can formally study 
how the 
Pohlmeyer reduction should be implemented in this case.
 Earlier discussions of this  pure $AdS_n$  reduction 
 appeared in \ci{deve,jevi} and we shall explain their relation to our approach. 



Let us start with the simplest case of $AdS_2=F/G=SO(2,1)/SO(1,1)$  and 
 use the standard matrix representation for $SO(2,1)$ by $3\times 3$ orthogonal
 matrices   with 
the  subgroup $SO(1,1)$  embedded diagonally (the signature choice is   $(--+)$).
 The Lie algebras are
denoted by $\algf=so(2,1)$ and 
$\algg=so(1,1)$. The orthogonal decomposition $\algf=\algp\oplus \algg$ 
induces the decomposition
$J=P+\A$ of the $\algf$-current $J=f^{-1}df$, \ $f \in F=SO(2,1)$.
The Virasoro constraints
\begin{equation}
 \tr(P_+P_+)=\tr(P_-P_-)=0\,
\end{equation}
imply that $P_\pm$ are proportional to $T_+$ or to  $T_-$ given by
\begin{equation}
\label{Tpm-ads2}
T_+=\bp 
 0&1&1\\
-1&0&0\\
1&0&0
\ep
\,, \qquad
T_-=\bp
 0&1&-1\\
-1&0&0\\
-1&0&0
\ep\ . 
\end{equation}
Note that these two choices are gauge inequivalent,  i.e.
 $T_+\neq g^{-1}T_-g$ for any $g \in G$.\foot{However, if one replaces $SO(2,1)/SO(1,1)$ 
 with the coset of slightly larger groups, namely, $O(2,1)/O(1,1)$
then there exists such $g$ that $T_+=g^{-1}T_-g$, e.g., $g=diag(1,1,-1)$ with 
$\det g =-1$ so that $g$ does not belong to $SO(2,1)$.}

Let us now consider   two options: 
(i) both $P_+$ and $P_-$ 
 are proportional to $T_+$ (or $T_-$);
  (ii) $P_+$ is proportional to $T_+$ and $P_-$  -- to $T_-$.
In the first case 
 the dynamics is trivial.  
Indeed, the  $\algg$-component of the MC equation takes the form
 $\d_-\A_+-\d_+\A_-=0$
($\commut{P_-}{P_+}$ vanishes due to the assumption
 that both  components are proportional to $T_+$).
This implies that $\A_\pm$ can be set to zero by a gauge transformation. 
The remaining equations of motion take the form
 $\d_-P_+=0, \ \  \d_+P_-=0$
and can be satisfied by making  appropriate conformal transformations.

In the second case
\be 
 P_+=p_+ T_+  \ , \ \ \ \ \ \  \ \ \    \ P_-=p_- T_-  \ , \ee
  and by a  gauge transformation
one can set $p_+=m=$const.
 Parametrizing $P_-=p_- T$ as $P_-=m e^{2\phi} T_-$ 
where $\phi$ is a new field we find that 
 the   Virasoro
constraints and part of {the} equations of motion are thus solved by 
\begin{equation}
P_+ = mT_+\ , \quad P_-=me^{2\phi} T_- \,, \quad \A_+=-\d_+\phi\ R_1\,, \quad \A_-=0\,,\ \ \ \ \
R_1=\bp
 0&0&0\\
0&0&1\\
0&1&0
\ep \,,
\end{equation}
where $R_1$ is an 
element of $\algg$.
Note that $\commut{R_1}{T_\pm}=\mp T_\pm$ and
$\commut{T_-}{T_+}=-2R_1$, i.e. $R_1,T_\pm$ form the  $sl(2) \approx so(2,1)$ algebra.
The only remaining equation is  the $\algg$-component of the Maurer-Cartan one which  gives 
\begin{equation}
 \d_-\d_+\phi  + m^2 \,e^{2\phi}=0\,,
\end{equation}
i.e. the Liouville equation. 
It follows from 
\begin{equation}
\label{li}
 L= \d_+\phi \d_-\phi -  m^2  e^{2\phi}\,, 
\end{equation}
which is thus the Lagrangian of the corresponding reduced theory.
Note that $m$ here can be set to any fixed value  by a shift of $\phi$
(the reduced theory has residual conformal invariance). 

The point we would like to make is that this model  can be viewed as a limit 
 of the Pohlmeyer-reduced model for  strings on $AdS_2\times S^1$ of the type 
 discussed in the previous sections.
Indeed, in this case  choosing the conformal gauge and fixing the residual
conformal freedom by choosing the angle of $S^1$ as $\psi= \mu \tau$ 
 the reduced theory is described by the sinh-Gordon Lagrangian 
\begin{equation}
L =\d_+\varphi \d_-\varphi-\frac{\mu^2}{2}\cosh 2\varphi  \ . 
\end{equation}
Introducing  $\phi=\varphi+\ln \mu$ we get 
\begin{equation}
L =\d_+\phi   \d_-\phi-\frac{\mu^2}{4}
(\mu^{-2} e^{2\phi}+\mu^2 e^{-2\phi} )\,.
\end{equation}
Then taking the limit $\mu \to 0$  we get   precisely the 
Liouville Lagrangian \eqref{li}  (with $m= \ha$).
This is just a manifestation of the 
fact  that   solutions where string  moves only in $AdS_2$ 
can be obtained  as a limit of solutions where it moves also along $S^1$.

Starting with string theory on 3-dimensional space  $AdS_2\times S^1$
one finds  the reduced Lagrangian  by completely fixing the reparametrization freedom and it thus
  contains just 3-2=1  physical degree of freedom. 
  At the same time, while  string theory on $AdS_2$ should have  no dynamical (transverse) degrees
   of freedom,  this is an apparent contradiction with 
 the reduced Lagrangian  \rf{li} depending on one field $\phi$. The resolution of this puzzle 
 is that the corresponding Liouville action  is still invariant under the conformal diffeomorphisms 
 which in present case are remnants of the original  reparametrization freedom  and should thus
  be treated as a gauge symmetry. 
 Fixing this symmetry  should leave no dynamical degrees of freedom.

Analogous considerations can be also applied to the reduced model for strings on $AdS_3$.
Starting from the reduced model for strings on $AdS_3\times S^1$ described by the Lagrangian
\begin{equation}
L= \d_+ \phi \d_- \phi   +   \tanh^2 \phi\  \d_+ \theta \d_- \theta- \frac{\mu^2}{2} \cosh 2\phi \ , 
\end{equation}
 the equations of motion  are 
\begin{align}
\d_+ \d_- \phi   -  \frac{\sinh\phi}{\cosh^3\phi}  \d_+ \theta \d_- \theta
 + \half \mu^2 \sinh2 \phi &=0\ , \\
\d_+(\tanh^2\phi\ \d_-\theta)+ \d_-(\tanh^2\phi\ \d_+\theta)&=0\,.
\end{align}
Writing them in  terms of the  rescaled variables
 $\phi^\prime=\phi+\log \m$ and $\theta^\prime=2\sqrt{2}\mu\theta$
and taking  the limit $\mu\to 0$ we get 
\begin{align}
\d_+ \d_- \phi^\prime   +  \half e^{2\phi^\prime}\d_-\theta^\prime \d_+ \theta^\prime
 - \half e^{-2 \phi^\prime} =0\ , \ \ \ \ \ \ \ \ \ 
\d_+(\d_-\theta')+ \d_-(\d_+\theta')=0\,.
\end{align}
The second equation can be  solved as $\theta'=\zeta_+(\sigma^+)+\zeta_-(\sigma^-)$. 
In terms of 
$\phi=\phi^\prime-\frac{1}{4}\ln{(\d_+\theta'\d_-\theta')}$ the first equation takes the form
$
\d_+ \d_- \phi   +  \sqrt{\d_+\zeta_+\d_-\zeta_-}\sinh 2 \phi=0, 
$
which can be put  into a simpler sinh-Gordon form 
\be
\d_+ \d_- \phi   +  \sinh{2\phi}=0\,
\ee
 by a $\zeta_\pm$-dependent 
conformal reparametrization of the 
 worldsheet coordinates.
 This then agrees with the result of the earlier discussion ~\cite{deve,jevi} 
 of the Pohlmeyer reduction of the $AdS_3$ sigma model (starting with the equations of motion in the 
 formulation in terms of  embedding coordinates).
Note that the $\mu\to 0$ limit of the $AdS_3 \times S^1$  theory we have used 
was taken  at the level of the equations of motion. 
It  cannot be  directly implemented 
at the Lagrangian level starting with 
the  Lagrangian of the hyperbolic
  CSG model (the reduced model for strings on $AdS_3 \times S^1$)    but it may be possible to
  take it 
  at the level of the  extended gWZW  action containing additional  
  gauge fields.\foot{To  
  get a smooth limit at the action level one should presumably 
incorporate more fields, going back to the gWZW formulation of the reduced 
theory for $AdS_3 \times S^1$.}

\bigskip

Let us now comment on  the general case of the  coset $F/{G}=SO(2,n-1)/SO(1,n-1)$. We shall  
use the standard 
matrix representation and assume that  the signature   is   $(--+\cdots +)$. 
The subspace $\algp=\algf\ominus\algg$
is then represented by elements with nonvanishing first raw and first column. 
Let $e_i$  ($i=0,\ldots,n-1$)
be the standard orthonormal basis in $\algp\ominus \algg$ with $\tr(e_0e_0)=1$
 and $\tr(e_ie_i)=-1$ for $i>0$.  The current components
$P_\pm$ then decompose as $P_\pm=P_\pm^i e_i$.
By making a $G$-gauge transformation one  can always satisfy the Virasoro constraint $\tr(P_+P_+)=0$
by (here we set an arbitrary mass scale $m$ that one can put in front of $T$  to 1) 
\begin{equation}
P_+= T, \qquad\quad T=
\bp
 0&1&1&0&\cdots&0\\
-1&0&0&0& \cdots&0\\
1&0&0&0& \cdots&0\\
0&0&0&0& \cdots&0\\
\cdots&\cdots&\cdots&\cdots& \cdots&\cdots\\
0&0&0&0& \cdots&0
\ep
\,.
 \end{equation}
Here $T$ is an obvious generalization of 
$T_+$ in the  $AdS_2$ case (see \eqref{Tpm-ads2}), but 
 unlike the $AdS_2$ case  in higher dimensions there are no inequivalent choices for 
$T$: analogs of  $T_+$ and $T_-$ are related by  gauge transformations. 

In the gauge where $P_+= T$ the equations of motion $\d_-P_++\commut{\A_-}{P_+}=0$ can be 
solved for $\A_-$ as
$\A_-=A_-$, where $A_-$ is an arbitrary $\algh$ valued field, with $\algh\subset\algg$ 
being a centralizer of $T$ in $\algg$.
Let $G_0\simeq SO(n-1)$ be diagonally embedded into $G$. 
By $G_0$ transformation one can always set 
$P^i_-=0$ for $i>1$. The Virasoro constraint $\tr(P_-P_-)=0$
then implies $(P_-^0)^2-(P_-^1)^2=0$, i.e.
 $P_-^0=\pm P_-^1$ (and by specializing the $G_0$ transformation one can also
set $P_-^1=P_-^0$). This allows one to use the following parametrization
\begin{equation}
 P_-=e^\phi g^{-1}T g\ , \ \ \ \   \ \ \ \ \ \  g \in G_0 \ , 
\end{equation}
where $g$ and $\phi$ are the new variables.\foot{In the $AdS_2$ case $G_0$ was 
 trivial so that the field $g$
was  not present.}
 The general solution to the  equation  $\d_+P_-+\commut{\A_+}{P_-}=0$,
  considered as a condition  on $\A_+$, is
\begin{equation}
\A_+=g^{-1}\d_+g + g^{-1}A_+ g -\d_+\phi\  g^{-1}{R}_1 g \ , \ \ \ \ \ \  g \in G_0 \ , 
\end{equation}
where  ${ R}_1$ is a  basic element of the subgroup (which is an obvious generalization
of $R_1$ in the $AdS_2$ case).
The only remaining equation is the $\algg$-component of the MC equation  that gives
\begin{equation}
\label{MC-adsn}
 D_-(g^{-1}\d_+g + g^{-1}A_+ g -\d_+\phi\ g^{-1}R_1 g)-D_+A_-=-e^{\phi}\commut{g^{-1}Tg}{T} \ . 
\end{equation}
Note that contrary to the standard Pohlmeyer reduction 
here we did  not fix  the residual conformal symmetry:  
eq. \eqref{MC-adsn}
is conformally  invariant with $g$ transforming as a scalar 
and $\phi$ as a Liouville field.

It remains  to be understood in general  how to find a Lagrangian from which \eqref{MC-adsn} 
may follow. For that we  may need to  fix the residual conformal symmetry; 
that may also help to explain the relation to  other reduced Lagrangians in the
literature \ci{deve,jevi}. For example, in the  $AdS_3$ case 
 we are left with 
 two  independent fields $\phi$ and $g$  while in \ci{deve,jevi}
 one finds  $\del_+ \del_- \a   - e^\a - u v e^{-\a} =0, \ 
 \del_+ u=0, \del_- v =0$  and after  solving for $u,v$ 
 (which essentially fixes the conformal symmetry) and redefining $\a$ one ends up with 
 sinh-Gordon equation  for a single  dynamical field. 
The analogous step of solving   for $g$  in $AdS_3$ example can be done in our case
(cf. the above discussion) 
but generalization  to  higher dimensional cases remains to be worked out. 

Let us stress again that in the context  of the \adss theory it is natural to view the 
subsector  of the pure $AdS_5$ solutions as a $\mu\to 0$ limit of the general 
string motions as described by the reduced theory of \ci{GT}.




\subsection{The vacuum and perturbative expansion of  the reduced model}

Going back to the reduced  Lagrangian of a
  $F/G$ bosonic coset model  which is similar to \rf{gw}
one needs to choose an  $H$-gauge to isolate the physical degrees of freedom. 
One option is to impose  $A_+=A_-=0$ which is possible at the level of  the equations of motion  \ci{GT}.
In this  gauge  one gets the equation of non-abelian Toda theory 
 with $g=\id$ 
as a natural vacuum point. The expansion near this point leads to  massive excitation spectrum 
with $\mu$ playing the role of a fiducial mass scale  which appears 
due to  spontaneous breaking of the
residual conformal invariance  by the condition like $t=\mu \tau$ in the $R_t \times F/G$ case. 
A drawback of this  ``on-shell'' approach 
is that  the equations in the $A_+=A_-=0$  gauge do not in general 
 follow from a local Lagrangian for the remaining independent degrees of freedom.

If instead one imposes the gauge on the group element $g$
and then integrates out $A_{\pm}$ one,  in general, 
 gets a sigma model with target space  metric which is singular at the natural vacuum point 
$g=\id $, so that the  perturbative expansion near this point appears to be  not well defined.
 This is due to the fact that the term $A_+A_--g^{-1}A_+g A_-$ in 
the vector-gauged WZW  model  is degenerate
at $g=\id $. In the case when $\algh$ is Abelian this problem 
can be cured by using the automorphism
$\tau(A)=-A$ as we did in the $AdS_3\times S^3$ case.
 However, already for $S^n$ or $AdS_n$
with $n\geq 4$ the  gauge algebra $\algh$ is nonabelian 
and this modification does not help.

One may  try  a more general modification of the Lagrangian as in the asymmetrically gauged WZW model
\cite{qu}. Namely, one  may choose  the gauge groups acting 
from the left and the right to be different embeddings of $H$ into $G$.
However,  this generalisation does not seem to be  relevant in the Pohlmeyer reduction context 
as
the left and the right gauge groups are determined by the choice of the 
fixed elements $T_+$ and $T_-$ from
$\algp=\algf\ominus\algg$  which define   $P_+=\mu T_+ $ and $P_-= \mu g^{-1} T_- g$
that solve the Virasoro conditions. In fact, for a rank 1 coset 
 all such  choices are equivalent and,  moreover,  $T_-$ can
be made equal to $T_+$ by an  appropriate redefinition of the field $g$.

An  alternative to the ``on-shell'' gauge on $A_\pm$ or the ``off-shell'' gauge on $g$ 
is an intermediate choice: to treat $g$ and $A_\pm$ on an equal footing, 
expand near $g=\id, \ A_\pm =0$   point and  impose a gauge on some combination of fluctuations 
of $g$ and $A_\pm$. That may lead to a non-degenerate perturbation theory
but it is not clear  a priori  if all of the resulting modes are then massive. 
A closely related possibility is to parametrize $A_+ = h^{-1} \del_+  h, \ 
A_- = h'^{-1} \del_-  h'$ and  then replace the gWZW part of the action by a 
difference of the two WZW actions 
$I( h^{-1} g h') - I( h^{-1}  h')$. One would then need to decide how 
gauge-fix (and redefine) the fields $g,h,h'$ to make  the expansion near $g=\id$ regular.
Further discussion of this  will appear in  \ci{RT}.

More generally, one  may   consider an  expansion near a non-trivial background 
of the reduced model that
corresponds to some solitonic solution of the original string model. 
For example, a  general  constant solution of the 
complex sine-Gordon theory corresponds to a rigid string solution on $R_t \times S^3$ 
and expanding near it leads to a non-degenerate (and UV finite) perturbation theory \ci{RT}. 
More generally, vacuum 
solutions with  constant Lagrange multiplier
for the embedding coordinates 
(or constant  value of the field that enters the potential of the reduced model 
 for strings on 
$R \times  S^n$  or  $AdS_n \times  S^1$  or $AdS_n \times S^n$)
correspond to rigid  circular strings 
with several angular momenta   constructed in  \ci{art}.\foot{An example is $(E,S;J)$   circular 
 string: it is stretched along a circle 
in $AdS_3$  and a circle in $S^1$. It has as its  charges  the energy  $E$  and  the spin $S$  in $AdS_3$ 
and  the spin $J$ in $S^1$.  Less trivial   solitonic solutions of the reduced models correspond   to 
more complicated  ``inhomogeneous'' string solutions. One example is 
the ``giant magnon'' (on an infinite line) in $R \times S^2$ 
 that was  constructed in \ci{hm}  from the sine-Gordon  soliton (see also \ci{dor}).  
 It can be viewed \ci{mtt}  as a special   case   of 
 an infinite spin limit of  a folded $(J_1,J_2)$ string on 
 $R \times S^3$    where $J_1$ and $E$   are taken to infinity. 
 For regular closed string   the folded  $(J_1,J_2)$ string on 
 $R \times S^3$   \ci{ft3}  originates from a regular  soliton 
 of (complex)  sine-Gordon  model.  Same remark applies to folded  
 string  in $AdS_3 \times S^1$   \ci{ft2}  and spiky string \ci{kru} 
  that correspond to  solitons of the sinh-Gordon model \ci{jevi}.}

\subsection{Relation between solutions  of the reduced and  the original model}

Let us now discuss in which sense the classical dynamics of the 
reduced model  determines
the dynamics of the original theory for strings on $R_t \times F/G$. 
A natural dynamical variable of the original model
is a group element $f\in F$. The equations of motion and the constraints
 are expressed  in terms of the current
$J_\pm=f^{-1}\d_\pm f$ that automatically satisfies  the Maurer--Cartan equation. They read 
\begin{equation}
 D_+P_-=0\,,\quad D_-P_+=0\,, \qquad  -\ha \tr(P_+P_+)=\mu^2\,, \quad  -\ha\tr(P_-P_-)=\mu^2\,.
\end{equation}
The Pohlmeyer reduction procedure amounts  to imposing first a particular  $G$-gauge condition
(i.e. the \textit{reduction gauge}).\footnote{In the case of  strings on $AdS_n\times S^n$
one  needs  also  to use the conformal transformations in order to impose 
 $T_{++}=\pm \mu^2,\,\,T_{--}=\pm\mu^2$ in the $AdS_n$ or  $S^n$ sectors together 
 with  the Virasoro constraints.
 The same  also applies to the
Pohlmeyer reduction of the $F/G$ coset sigma model (in contrast to the reduction 
of strings on $F/G\times R_t$
where $T_{++}=\mu^2,\,\,T_{--}=\mu^2$ are just the Virasoro constraints
 in the conformal   gauge supplemented   by the condition $t=\mu \tau$
 fixing the residual conformal diffeomorphisms).}
In this gauge the components of the current $J$ are expressed in terms of
the new fields $g,A_\pm$ satisfying the equations of motion of the reduced system
\begin{equation}
 J_+=g^{-1}A_+g+g^{-1}\d_+g+\mu T\,,\qquad J_-=A_-+\mu g^{-1}Tg\,.
\end{equation}
Since  the equations of motion of the reduced system are essentially 
 the MC equations for $J$ (parametrized by $g,A_\pm$)
one can reconstruct the configuration $f(\sigma^+,\sigma^-)$ of the original model 
 in terms
of a solution $ (g,A_\pm)$ of the reduced system or  $J_\pm(\sigma^+,\sigma^-)$ 
 by solving the auxiliary linear problem:
\begin{equation}\label{recovery}
 f^{-1}\d_+f=g^{-1}A_+g+g^{-1}\d_+g+\mu T\,,\qquad f^{-1}\d_-f= A_-+\mu g^{-1} Tg\,.
\end{equation}
This system has a unique solution for any initial data 
$f|_{_{\sigma^\pm=\sigma^\pm_0}}=f_0$ \ ($f_0\in F$) specified at
a given point on the world sheet.

The group $F$ naturally acts from the left on the initial data. This action
induces the left action of $F$ on the space of solutions to~\eqref{recovery}.
In this way one recovers the  global $F$  symmetry present in the original model
but not seen   in the reduced model, i.e.  
in the formulation in terms of the  currents
 (which are the invariants of the global left $F$-action).
To summarize, any solution to the original system is equivalent to a
 solution of \eqref{recovery}
for an  appropriate choice of the solution $J_\pm(\sigma^+, \sigma^-)$ of the reduced system and an 
initial condition $f_0$. 
The equivalence means that they are related by a $G$-gauge 
transformation and a conformal  reparametrization.

\subsection{Conserved charges }

While the original global $F$ symmetry is not visible in the reduced model 
formulated in terms of the currents, one can still classify the solutions (and thus states)  
of the reduced model by values of (higher) Casimir operators which are also invariant under $F$. 

Indeed, let  us consider  the counterpart of the natural vacuum solution 
of the reduced system $g=\id,\ A_\pm=0$ in the original string  model on $R_t \times F/G$. 
Here we shall use   the on-shell gauge $A_\pm=0$.
Then eq.~\eqref{recovery} takes the form
\begin{equation}
 f^{-1}\d_+f=\mu T\,,\qquad f^{-1}\d_-f =\mu T\,, 
\end{equation}
and it can be formally  solved by
\begin{equation}
f=  e^{\mu(\sigma^++\sigma^-)T}\ f_0=e^{\mu\tau T}\ f_0 \ . 
\end{equation}
The $\algf$-valued conserved (Noether) current corresponding to the global left 
action of the group  $F$ on the coset $F/G$  has the form
(see,  e.g., \cite{ef})
\begin{equation}
 j_a=f\ (f^{-1}\d_a f)_\algp f^{-1} \ ,  
\end{equation}
as one can see from the fact that  the equations of motion for the coset model can be written as
$\d_a j^a=0$.
Evaluating the corresponding  conserved charge 
on the above solution  one gets:
\begin{equation}
M=\int d\sigma\  j_\tau= \mu \int d\sigma\  f_0 T f_0^{-1}\,,
\end{equation}
where we have used that  $j_\tau=(f^{-1}\del_\tau  f )_\algp=T$.
Assuming the space direction $\s$ to be compact ($0 < \s \leq 2\pi$) 
one gets a non-zero value  for the quadratic Casimir
\begin{equation}
 K\equiv -\half\tr(MM)=(2\pi \mu)^2  \ . 
\end{equation}
Here we have used the convention $\tr(TT)=-2$.

For more general solutions it is nontrivial to find the explicit 
values  of the Casimir operators. 
For example, 
let us consider the    $S^n$ model described by the embedding coordinates: \  
$
L = \del_+ X_i \del_- X_i, \ \    X^2=1 $. Then the 
$F=SO(n+1)$   symmetry  leads to the Noether 
currents conserved on the equations of motion 
\be 
(j_{ik})_a=  X_i \d_a X_k - X_k \d_a  X_i\,, \qquad \d_a j_{ik}^a=0\ . 
\ee
 The corresponding charges  and the quadratic Casimir of $SO(n+1) $
\be 
M_{ik} = \int  d \s \  (j_{ik})_\tau \ , \ \ \ \ \ \ \ \ 
K=\half  M_{ik} M_{ik}
  \ee 
are then conserved in $\tau$.
Explicitly, 
$$
(j_{ik})_\tau (j_{ij})_\tau  = 2  \d_\tau  X_i \d_\tau X_i = - 2 \tr ( P_\tau  P_\tau )
$$
Since in the  vacuum  of  the reduced model  $g=1$ we have 
$
P_+ = \mu T, \ \ 
         P_-=\mu g^{-1} T g = \mu  T
$
then  $P_\tau  = P_+  +  P_- = 2 \mu T$, \  $ P_\sigma  =0$
so  that again 
$ (j_{ik})_\tau (j_{ij})_\tau= 8 \mu^2$.
But in general 
\be K = \half  \int d\s   (j_{ik})_\tau (\s)   \int d\s'   (j_{ik})_\tau (\s')   \ee
so that  it is not clear if $K$ is non-zero and is related to $\mu$
unless $ (j_{ik})_\tau$ is constant in $\s$.
The reduced 
 theory thus does not tell us 
 much about the charges of the original theory  before  we actually solve
the linear problem for $X_i$  or, equivalently,  for $f$.

\bigskip

{\bf Acknowledgments}

\noindent
We are grateful to G. Arutyunov,  U. Lindstrom, L. Miramontes, 
G. Papadopoulos and especially  R. Roiban for useful discussions. 
A.A.T. is grateful to the organizers of the International conference
 on progress of string theory and quantum field theory
 at the Osaka City University in December 2007 
for their kind hospitality. 
The work of M.G. was partially supported by the Dynasty foundation,
RFBR grant 07-01-00523, and grant LSS-4401.2006.2.

\bigskip

\appendix

\section{Details of  computation of $AdS_3 \times S^3$  reduced Lagrangian}\label{app}

We shall  use the following basis in $su(1,1)$ and $su(2)$ 
\begin{align}
\bar R_1&=
\left(\begin{array}{cc}
0&1\\
1&0\\
\end{array}
\right)\,,
&\qquad
\bar R_2&=
\left(\begin{array}{cc}
i&0\\
0&-i\\
\end{array}
\right)\,,
&\qquad
\bar R_3&=
\left(\begin{array}{cc}
0&i\\
-i&0\\
\end{array}
\right)\,,\\
R_1&=
\left(\begin{array}{cc}
0&\imath\\
\imath &0\\
\end{array}
\right)\,,
& \qquad
R_2&=
\left(\begin{array}{cc}
i&0\\
0&-i\\
\end{array}
\right)\,,
&\qquad
R_3&=
\left(\begin{array}{cc}
0&-1\\
1&0\\
\end{array}
\right)\,.
\end{align}
The parametrization of $g$ in terms of the Euler angles reads 
\begin{align}
g_A=(g_A)_2(g_A)_1(g_A)_2\,,\qquad (g_A)_1=\exp{(\phi \bar R_1)}\,,\quad (g_A)_2=\exp{(\half \chi \bar R_2)}\,,\\
g_S=(g_S)_2(g_S)_1(g_S)_2\,,\qquad (g_S)_1=\exp{(\varphi  R_1)}\,,\quad (g_S)_2=\exp{(\half \theta R_2)}\,,
\end{align}
or more  explicitly:
\begin{align}
(g_A)_1&=
\left(\begin{array}{cc}
\cosh \phi &\sinh \phi\\
\sinh \phi & \cosh \phi
\end{array}
\right)\,,&\qquad
(g_A)_2&=
\left(\begin{array}{cc}
\exp{(\half \imath \chi)} &0\\
0& \exp{(-\half \imath \chi)}	
\end{array}
\right)\,,\\
(g_S)_1&=
\left(\begin{array}{cc}
\cos \varphi &\imath \sin \varphi\\
\imath \sin \varphi & \cos \varphi
\end{array}
\right)\,,&\qquad
(g_S)_2&=
\left(\begin{array}{cc}
\exp{(\half \imath \theta)} &0\\
0& \exp{(-\half \imath \theta)}	
\end{array}
\right)\,.
\end{align}
In order to solve \eqref{A-const-v} \eqref{hatA-v} for $A_+,A_-$, let us note the following
useful relations:
\begin{align}
(g_A^{-1}\d_+ g_A )_\algh & =\d_+\chi\frac{1+\cosh2\phi}{2}\bar R_2\,,&\qquad
 (g_A^{-1}A^A_+g_A)_\algh & =\cosh 2\phi\ A_+^A\,,\\
(g_S^{-1}\d_+ g_S )_\algh & =\d_+\theta\frac{1+\cos 2\varphi}{2}R_2\,,&\qquad 
(g_S^{-1}A^S_+g_S)_\algh & =\cos 2\varphi\ A_-^S\,,
\end{align}
where we have used
\begin{align*}
(g_A)_1^{-1}\bar R_2(g_A)_1&=\cosh 2\phi\ \bar R_2+\sinh 2\phi\ \bar R_3\,,&\qquad 
(g_A)_2^{-1}\bar R_1(g_A)_2&=\cos \chi\ \bar R_1-\sin\chi\ \bar R_3\\
(g_S)_1^{-1}R_2(g_S)_1&=\cos 2\varphi\  R_2+\sin 2\varphi\ R_3\,,&\qquad 
(g_S)_2^{-1}R_1(g_S)_2&=\cos \theta\ R_1-\sin \theta\  R_3\,,
\end{align*}
\begin{align*}
(g_A)_2^{-1}(g_A)_1^{-1}\bar R_2(g_A)_1 (g_A)_2&=\cosh 2\phi\ \bar R_2+\sinh 2\phi\ 
\sin \chi\ \bar R_1+\sinh 2\phi\ \cos \chi\ \bar R_3\,,\\
(g_S)_2^{-1}(g_S)_1^{-1}R_2(g_S)_1 (g_S)_2&=\cos 2\varphi\ R_2+\sin 2\varphi\ 
\sin\theta\ R_1+\sin 2\varphi\ \cos \theta\ R_3\,.
\end{align*}
Parametrizing the fermions according to~\eqref{ferm-param} one  arrives at \eqref{ferm-term} 
and then gets  the explicit solution for $A_+,A_-$.

In computing the third line of \eqref{hi} the following relations are useful
\begin{equation}
 (\d_-g_Ag_A^{-1})_\algh=\frac{1+\cosh 2\phi}{2}\d_-\chi\bar R_2\,,\qquad
 (\d_-g_Sg_S^{-1})_\algh=\frac{1+\cos 2\varphi }{2}\d_-\theta R_2\,.
\end{equation}
Let us  note  also  that the fermionic interaction term 
entering the Lagrangian 
can be computed in terms
 of $2\times 2$ matrices using 
the observation
that the two contributions to the supertrace of $g^{-1}\Psi_2 g\Psi_1$ are
 complex conjugates of one another
so that 
$\str(g^{-1}\Psl^\prime g\Psr^\prime)
=
-2 {\rm Im}[\tr(g_A^{-1}\psi_2 g_S \psi_1^\dagger\Sigma)]$.

\bigskip

Let us also give some details on direct computation of the reduced Lagrangian in the axial gauging case.
In terms of the Euler angles the parametrization of the group element reads as 
\begin{equation}
g_A=
\left(\begin{array}{cc}
 \cosh \phi & e^{-\imath\chi}\sinh \phi\\
e^{\imath\chi}\sinh \phi &\cosh \phi
\end{array}
\right)\,,\qquad
g_S=
\left(\begin{array}{cc}
 \cos \varphi &\imath e^{-\imath \theta}\sin \varphi\\
\imath e^{\imath \theta}\sin \varphi & \cos \varphi
\end{array}
\right)\,.
\end{equation}
One  can then solve for the gauge fields  using their equations 
\bea
\label{A-const-ax}
&& -A_+=(\hat A_+)_\algh\,,  \qquad \hat A_+ \equiv  g^{-1}\d_+g+g^{-1}A_+g 
-\half\commut{\commut{t}{\Psr^\prime}}{\Psr^\prime}\,,
 \\
 \label{hatA-ax}
&&  A_-=(\hat A_-)_\algh \,, \qquad
\hat A_- \equiv  g\d_-g^{-1}-gA_-g^{-1} 
-\half\commut{\commut{t}{\Psl^\prime}}{\Psl^\prime}\,.
\eea
following from the Lagrangian \eqref{L-tot} with $\tau(a)=-a$.
Similarly to the previous  case  one finds
\bea
A^S_+=-\frac{\d_+\theta(1-\cos2\varphi)+2(\alpha\beta-\gamma\delta)}{2(1+\cos{2\vp})}R_2\,,\quad
A^A_+=-\frac{\d_+\chi(1-\cosh2\phi)-2(\alpha\beta-\gamma\delta)}{2(1+\cosh{2\phi})}\bar R_2\,, \no
\eea
and also the  expressions  for $A_-$
with  $\d_+\chi \to -\d_-\chi$ and $\alpha\beta-\gamma\delta   \to    \l \nu - \rho \sigma $.
Note also the following useful relations:
\begin{equation}
 (\d_-g_Ag_A^{-1})_\algh=-\frac{1-\cosh 2\phi }{2}\d_-\chi\bar R_2\,,\qquad
 (\d_-g_Sg_S^{-1})_\algh=-\frac{1-\cos 2\varphi }{2}\d_-\theta R_2\,.
\end{equation}
Eliminating $A_\pm$ as in the vector gauge case
one finds the bosonic part of  the  Lagrangian that comes from the WZW  and potential
terms
\bea 
&&L^{axial}_1= 
   \dpp \varphi \dmm \varphi  +   \ha (1 - \cos 2  \varphi ) \ \dpp \theta \dmm \theta \no \\
&&   +
   \dpp \phi \dmm \phi       - \ha ( 1 - \cosh 2   \phi ) \ \dpp \chi  \dmm \chi
   + \frac{\m^2}{2} (\cos 2\varphi  - \cosh 2\phi) \ . 
\eea
The  fermionic interaction term together with the fermionic kinetic terms give 
\bea
&&L^{axial}_{2} = 
\alpha\d_-\alpha+\beta\d_-\beta+\gamma\d_-\gamma+\delta\d_-\delta+\l\d_+\l+\nu\d_+\nu
+\rho\d_+\rho+\sigma\d_+\sigma\no  \\
&&-2\mu\Big(
\cosh{\phi}\cos{\varphi}(-\l \beta+\nu\alpha+\rho\delta-\sigma\gamma) 
+\sinh{\phi}\sin{\varphi} \big[ \cos{(\chi+\theta)}(-\rho\alpha-\sigma\beta\no\\
&& +\l\gamma+\nu\delta) -\sin{(\chi+\theta)}(-\rho\beta+\sigma\alpha-\lambda\delta+\nu\gamma)\big]\Big) \,.
\eea
Finally, the terms that originate from the elimination of $A_\pm$ 
are 
\bea
&&L^{axial}_{3} = 
  -\frac{[\d_+\chi\ (1-\cosh 2\phi)\ -2 (\alpha\beta-\gamma\delta)]
 [\d_-\chi\ (1-\cosh 2\phi)\ -2(\l\nu-\rho\sigma)]}{2(1+\cosh 2\phi )} \no  \\
&&+  \frac{[\d_+\theta\ (1-\cos 2\varphi)\ +2(\alpha\beta-\gamma\delta)][\d_-
 \theta\ (1-\cos 2\varphi)\ +2(\l\nu-\rho\sigma)]}{2(1+\cos 2\varphi) }\ . 
\eea
Then 
$L_{tot}^{axial}= L^{axial}_1 + L^{axial}_2 + L^{axial}_3$.
The purely bosonic terms in $L^{axial}_1 $ and $L^{axial}_3$ combine into the direct sum 
of the CSG   action and its ``hyperbolic'' counterpart
\begin{equation}
   L^{axial}_B =
   \dpp \varphi \dmm \varphi  +  \tan^2 { \varphi }\ \dpp \theta \dmm \theta
   +
   \dpp \phi \dmm \phi  +   \tanh^2 { \phi }\ \dpp \chi  \dmm \chi
    +\frac{\m^2}{2} (\cos 2\varphi  - \cosh 2\phi) \,,
\end{equation}
while the fermionic terms give 
\begin{multline}
L^{axial}_{F}= 
L^{axial}_2+\tan^2(\varphi)[\d_+\theta(\lambda\nu-\rho\sigma)+\d_-\theta(\alpha\beta-\gamma\delta)]\\
-\tanh^2(\phi)[\d_+\chi(\lambda\nu-\rho\sigma)+\d_-\chi(\alpha\beta-\gamma\delta)]
+(\alpha\beta-\gamma\delta)(\lambda\nu-\rho\sigma)[\frac{1}{\cos^2\varphi}-\frac{1}{\cosh^2\phi}]\,.
\end{multline}
Redefining the fermions according to $\alpha\to\delta,\ \delta\to-\alpha,\ \beta\to-\gamma,\ 
\gamma\to\beta$
and combining all of the terms together one indeed gets Lagrangian~\eqref{L-tau}.

\end{document}